\documentclass[a4paper,fleqn]{cas-dc}

\usepackage{natbib}

\def\tsc#1{\csdef{#1}{\textsc{\lowercase{#1}}\xspace}}
\tsc{WGM}
\tsc{QE}
\tsc{EP}
\tsc{PMS}
\tsc{BEC}
\tsc{DE}

\begin{document}
\let\WriteBookmarks\relax
\def\floatpagepagefraction{1}
\def\textpagefraction{.001}
\shorttitle{}
\shortauthors{J}

\title [mode = title]{Unveiling the Cosmic Dawn with SHARP: 
Probing extended Lyman-$\alpha$ nebulae 
in a Universe less than 600 Myr old}




\author[1]{S. Bisogni}[type=editor,
                        orcid=0000-0003-3746-4565]
\cormark[1]
\ead{susanna.bisogni@inaf.it}

\credit{Conceptualization, Methodology, Original draft preparation}

\affiliation[1]{organization={INAF-IASF Milano},
                addressline={Via Alfonso Corti 12}, 
                city={Milano},
                postcode={20133}, 
                country={Italy}}

\author[1]{G. Vietri}

\credit{Methodology, Review \& editing}

\author[2]{E. Piconcelli}

\credit{Methodology, Review \& editing}

\author[3]{F. Ricci}

\credit{Methodology, Review \& editing}

\author[4]{A. Travascio}

\credit{Methodology, Review \& editing}

\author[1]{P. Franzetti}

\credit{Software, Methodology}

\author[1]{A. Gargiulo}

\credit{Methodology, Review \& editing}

\author[1]{C. Mancini}

\credit{Methodology, Review \& editing}

\affiliation[2]{organization={INAF – Osservatorio Astronomico di Roma},
                addressline={Via Frascati 33},
                city={Monte Porzio Catone}, 
                postcode={00040},  
                country={Italy}}

\affiliation[3]{organization={Department of Mathematics and Physics, Roma Tre University},
                addressline={Via della
Vasca Navale 84},
                city={Rome},
                postcode={I-00146}, 
                country={Italy}}

\affiliation[4]{organization={INAF–OAT, Osservatorio Astronomico di Trieste},
                addressline={Via Tiepolo 11},
                city={Trieste},
                postcode={34131}, 
                country={Italy}}

\cortext[cor1]{Corresponding author}

\begin{abstract}
The existence of luminous quasars just a few hundred million years after the Big Bang challenges our understanding of both black hole growth and galaxy formation and evolution. These objects harbour supermassive black holes exceeding a billion solar masses (M$_{BH} > 10^{9} M_{\odot}$) by redshift $z\sim 6.5$-$7.5$, powered by extreme gas accretion. At the same time, their host galaxies are also undergoing intense star formation, consuming gas at the rate of hundreds of solar masses per year. Characterising the circumgalactic medium (CGM) and intergalactic medium (IGM) surrounding high-redshift quasars becomes an essential tool to understand the conditions that enable the rapid formation and evolution of these extreme sources. While in the last decades spatially resolved observations in the optical band have targeted CGM through Ly$\alpha$ nebulae surrounding $z \sim 2-6$ quasars, current instrumental limitations hamper observations of high-z ($z>8$) quasars that will be discovered by Euclid/Roman/LSST surveys. Despite the large fraction of neutral hydrogen at the epoch of reionisation, in the last decade several surprising Ly$\alpha$ detections have been obtained from sources deep in the epoch of reionisation. The unprecedented collecting area of ELT, coupled with the resolution and wavelength coverage of SHARP, specifically VESPER, will enable us to map for the first time $z>9$ Ly$\alpha$ emission down to the structures of size $\sim$150 pc, while simultaneously capturing their large-scale structure up to 100 kpc for the first time at this redshift. This will allow a major and long-awaited step forward in the exploration of quasars and galaxies formation and evolution deep in the epoch of reionisation.
\end{abstract}




\maketitle

\section{Introduction}
\label{introduction}

Scaling relations between supermassive black hole (SMBH) mass and bulge properties (e.g. velocity dispersion and stellar mass) in the local Universe \citep[see][]{Kormendy2013, McConnell2013, Heckman2014} suggest that host galaxies and their central AGN, despite the very different spatial scales involved, are closely linked and may have co-evolved over cosmic time.
With some notable differences, recent JWST observations suggest that this result may also hold at high redshift \citep[$z=4-7$, ][]{Pacucci2023}, implying that efficient energy coupling between SMBH and their large-scale environments was already in place in the early Universe.
However, confirming this scenario and uncovering the underlying physical mechanisms remain highly challenging, given the complex exchange of energy, momentum, and baryonic components between the SMBH, the host galaxy, and their surrounding environment. 

The Circumgalactic Medium (CGM), the diffuse, multi-phase gas within the virial radius of a galactic halo, plays a crucial role in SMBH-galaxy co-evolution, acting as the primary reservoir that replenishes the Inter Stellar Medium (ISM) and sustains both star formation and SMBH accretion. Understanding how different scales access this gas, and under what conditions it can cool and condense onto the ISM, is key to constraining the long-term growth of both the stellar and black hole components.
AGN feedback operates across scales ranging from the sub-parsec to the kiloparsec \citep[][ and reference therein]{Harrison2024}: while the SMBH is fed on the smallest scales, star formation occurs on galactic ones, and both processes depend on the availability and regulation of gas. It is therefore crucial to understand how feeding and feedback are balanced in terms of fuel supply.
The AGN can directly affect the ISM, regulating star formation and shaping galaxy-wide properties through the continuous injection of energy over time, rather than through isolated, short-lived episodes \citep[e.g. ][]{Piotrowska2022, Bluck2023, Terrazas2016}. At the same time, it can influence the CGM, thereby indirectly preventing or regulating star formation in the host galaxy \citep{Davies2020, Choi2018}.
The strength and impact of these processes depend on how efficiently AGN energy couples with the gas across different spatial and temporal scales. 

A key perspective on these mechanisms arises from the study of quasars at high redshift. To date, several hundred quasars have been discovered at $z>6$ \citep{Fan2023}, some extending the redshift frontier to
$z \sim 7.6$ \citep[e.g. ][]{Banados2018, Yang2020, Wang2021}. 
Remarkably, they host SMBH with $M_{BH} \sim 10^{9}-10^{10}\, M_{\odot}$ assembled within the first $\sim700$ Myr of cosmic time, efficiently growing mass onto smaller BH seeds \citep{Volonteri2021, Inayoshi2020}, fueled by the gas supplied by their host galaxies. At the same time, millimeter and submillimeter observations have shown that these hosts themselves are rapidly evolving, with star formation rates (SFRs) well above $100 M_{\odot} yr^{-1}$, sustained by the gas of the environment \citep{Decarli2018, Kim2019, Shao2019}.

To understand how these massive structures form and grow, it is crucial to investigate their environments. Several studies suggest that high-redshift quasars form within the most massive dark matter halos ($\gtrsim 10^{12} M_{\odot}$; e.g. \citealt{Costa2014, Volonteri2006}), and that sustaining their rapid growth requires a continuous inflow of fresh gas, either from the intergalactic medium (IGM) or through mergers with gas-rich halos (e.g. \citealt{Fumagalli2011, DiMatteo2012, Mayer2019}). Recent theoretical and simulation-based works \citep[e.g. ][]{Dekel2009, Nelson2013, Aung2024, Waterval2025} indicate that, at high redshift, gas can also accrete efficiently via cold, dense streams penetrating the halo, providing a steady fuel supply without the need for major mergers or extremely massive halos. These cold flows may therefore provide a viable mechanism sustaining both rapid black hole growth and intense star formation in the early Universe.

These scenarios can be tested through observations of gas reservoirs in the CGM, the region extending up to a few hundred kiloparsecs around high-redshift quasars.
At these early cosmic epochs, the cool gas in the CGM can be probed in emission. In particular, the intense UV radiation from the AGN can be reprocessed by the surrounding gas into Ly$\alpha$ photons, producing a diffuse, spatially extended glow of Ly$\alpha$ fluorescence \citep{Costa2022}. This mechanism is likely the main process that makes extended Ly$\alpha$ emission a powerful tracer of the distribution and kinematics of the CGM \citep{Cantalupo2005}.


\section{Narrow band imaging and long-slit spectroscopy of Ly$\alpha$ nebulae}

The first studies investigating Ly$\alpha$ emission in the CGM employed long-slit spectroscopy and narrow-band imaging of sources at intermediate redshifts \citep[e.g.,][]{ArrigoniBattaia2016, Roche2014}. These initial works revealed that Ly$\alpha$ nebulae with sizes of 10–50 kpc are nearly ubiquitous around intermediate redshift quasars. A subset of sources, typically located in galaxy overdensities, was found to host giant Ly$\alpha$ nebulae extending over 200 kpc or more, thus tracing a substantial fraction of the surrounding environment of the quasar.
A prominent example is the Slug Nebula \citep[$z\sim2.3$, ][]{Cantalupo2014}, the first giant Ly$\alpha$ nebula detected around a radio quiet quasar using the narrow-band technique. In this system, the Ly$\alpha$ emission arises either from the ionization of the nebular gas by the UV radiation of the quasar, followed by hydrogen recombinations, or from the scattering of Ly$\alpha$ photons originally produced in the quasar’s Broad Line Region. The Slug Nebula extends over a projected size of approximately 460 kpc, with a maximum observed surface brightness (SB) of $\sim 10^{-16}$ erg s$^{-1}$ cm$^{-2}$ arcsec$^{-2}$ and a total Ly$\alpha$ luminosity of $\sim 2 \times 10^{44}$ erg s$^{-1}$. These extreme values highlight the enormous energies involved in powering such systems.

However, both narrow-band imaging and long-slit spectroscopy face significant limitations, particularly for radio quiet quasars, which tend to host kinematically narrower nebulae than their radio-loud counterparts, where jets can contribute to broader and more complex kinematics. In narrow-band imaging, uncertainties in the quasar’s systemic redshift can cause the Ly$\alpha$ line to fall outside the filter bandpass, while long-slit spectroscopy suffers from slit losses due to the complex and often asymmetric morphology of the nebulae, which cannot be fully captured in a single slit position \citep[see ][]{Borisova2016}. In both cases, accurate PSF subtraction, crucial for isolating the faint extended emission, remains challenging.

\section{Integral field spectroscopy of Ly$\alpha$ nebulae}

Integral Field Spectroscopy (IFS) overcomes these limitations by enabling robust PSF subtraction, avoiding filter and slit losses, and, importantly, often providing large fields of view, which make it particularly well suited for mapping the diffuse and extended Ly$\alpha$ emission around quasars.

Thanks to the advent of IFU spectroscopy, and in particular the capabilities of MUSE, \cite{Borisova2016} obtained the first statistical sample of Ly$\alpha$ nebulae around radio quiet quasars.
Most existing statistical IFU surveys have focused on radio-quiet quasars, largely because they dominate optically selected samples and provide a baseline case in which the CGM is primarily illuminated by the quasar radiation field, without the additional mechanical impact of powerful radio jets. Extended Ly$\alpha$ nebulae are, however, also observed around radio-loud quasars, sometimes with enhanced surface brightness, anisotropic morphologies, and more complex kinematics that can be influenced by jet-CGM interactions \citep[e.g. ][]{Roche2014, Morais2017}. Throughout this paper we therefore specify radio-quiet versus radio-loud mainly to indicate whether jet-driven effects are expected to contribute, rather than because extended Ly$\alpha$ emission is exclusive to one class. Indeed, comparing radio-quiet and radio-loud systems is itself valuable for understanding baryon cycling, since episodic jet phases may leave long-lived imprints in the CGM \citep{Hardcastle2020}.
In \cite{Borisova2016}, giant nebulae with projected sizes exceeding 100 kpc were found around almost every radio quiet quasar at $z \sim 3-4$. However, this study also highlighted that the kinematic interpretation of the Ly$\alpha$ line remains highly uncertain\footnote{Only a few nebulae show resolved kinematic maps with a coherent, rotation-like pattern, for instance, the Enormous Lyman-alpha Nebula at $z \sim 3$ observed with MUSE by \cite{ArrigoniBattaia2018}, but such cases appear to be the exception rather than the rule.}. This is mainly due to the resonant nature of Ly$\alpha$, which leads to strong radiative transfer effects that make it extremely challenging, if not impossible, to directly recover the underlying gas kinematics. Simulations that implement Monte Carlo radiative transfer techniques therefore play a crucial role in disentangling the propagation of Ly$\alpha$ photons and in interpreting the observed velocity structures.

To date, several surveys have been carried out at different redshifts to investigate the CGM through its Ly$\alpha$ emission around quasars. Beyond the pioneering MUSE study of luminous quasars at $z \sim 3$–4 by \citet{Borisova2016}, notable examples include the KWCI survey at $z \sim 2$ \citep{Cai2019}, the MUSEUM survey at $z \sim 3$ \citep{ArrigoniBattaia2019}, the MAGG survey at $z \sim 3.5$ \citep{Fossati2021}, and the REQUIEM survey at $z \sim 6$ \citep{Farina2019}.
In some notable cases, extended emission from a metal-enriched CGM has also been detected around radio-quiet quasars in other lines, such as C\,\textsc{iv} and He\,\textsc{ii},\footnote{Both emissions offer insights into the interplay between the CGM and the central source. In particular, C\,\textsc{iv} emission suggests the presence of metal-enriched gas, plausibly linked to enrichment by quasar-driven outflows (see also Section~\ref{sec:future}); however, its surface brightness depends strongly on ionization parameter, gas density, and geometry/covering factor, so it is not a direct metallicity diagnostic on its own. Moreover, since C\,\textsc{iv} and He\,\textsc{ii} are not affected by resonant scattering, they can provide more direct constraints on the kinematics and physical conditions of the surrounding medium than Ly$\alpha$. The high surface brightness of extended Ly$\alpha$ emission, however, makes it the most widely studied tracer to date.} \citep[e.g.][ Travascio et al. (in prep.) for reporting on He\,\textsc{ii} nebulae in the MAGG and KCWI samples]{Guo2020, Travascio2020}. Such detections remain relatively scarce primarily because these lines are typically much fainter than Ly$\alpha$ and require very deep data and excellent PSF subtraction to reach the necessary surface-brightness limits.
Moreover, unlike absorption-line studies that can constrain CGM metallicities along individual sightlines, interpreting metal-line emission generally requires photoionization modelling and multiple line ratios, since the strength of C\,\textsc{iv} and He\,\textsc{ii} depends not only on metal abundance but also on ionization parameter, density, and geometry \citep[e.g.][]{ArrigoniBattaia2015}.

However, current observations are largely constrained by the wavelength coverage of optical instruments, which limits the accessible redshift range for studying Ly$\alpha$ emission. As emphasized by \citet{Farina2019}, pushing to higher redshifts is crucial: by $z \sim 4$, the Universe is already about 1.5 Gyr old, and a population of massive galaxies is already in place. To truly probe the earliest stages of galaxy formation, it is essential to extend these studies to even higher redshifts.
Current instrumentation limits such studies to $z \sim 7$. 
Recent JWST/NIRSpec observations have extended the study of Ly$\alpha$ nebulae to higher redshifts (see, for instance, in the next section, the nebula around GN-z11 at $z \sim 11$; \citealt{Scholtz2024}), but the relatively modest spectral resolution of these data ($R \sim 1000$) limits detailed investigations of the gas kinematics compared to current optical IFU such as MUSE or next-generation instruments like SHARP/VESPER ($R \sim 3000$).

\begin{table*}[t]
\centering

\begin{tabular}{l c c c c c c}
 \hline
 Instrument & $\lambda$ range & $R$ & Sampling & FoV & Multiplex & PSF FWHM$^\dagger$ \\
            & [$\mu$m]        &     & [arcsec] & [arcsec] & & [arcsec] \\
 \hline
 MUSE (VLT) & 0.465-0.93 & $\sim$1800-3600 & 0.20/pix & $60\times60$ & 1 & $\sim$0.5--0.8 \\
 KCWI (Keck)$^\ddagger$ & 0.35-1.08 & $\sim$1000-20000 & 0.35 to 1.4/pix & $20\times (8$ to $33)$ & 1 & $\sim$0.5-0.8 \\
 JWST/NIRSpec & 0.6-5.3 & $\sim$1000-2700 & 0.10/pix & $3\times3$ & 1 & $\sim$0.03-0.10 \\
 VESPER (ELT) & 1.2-2.4 & $\sim$3000 - 4000 & 0.031/pix & $1.7\times1.5$ (per IFU) & 12 & $\lesssim$0.05-0.10 \\
 \hline
\end{tabular}
\caption{Key instrumental parameters relevant for IFU studies of extended Ly$\alpha$ emission.
MUSE and KCWI enable systematic CGM studies at $z\sim2$--6 in the optical; JWST/NIRSpec extends the wavelength coverage to the reionisation era, often at more modest spectral resolution in current IFU programs; VESPER combines near-IR coverage (Ly$\alpha$ at $z>9$), $R\sim3000$, and AO-assisted spatial sampling, opening spatially resolved CGM studies in the first few hundred Myr.
$^\dagger$Indicative values: MUSE/KCWI are generally seeing-limited, JWST is diffraction-limited, and VESPER is expected to be AO-assisted (MCAO); actual performance depends on conditions and wavelength. $^\ddagger$KCWI parameters depend on the selected slicer/grating configuration; ranges are shown.}
\label{tab:ifu_compare}
\end{table*}

To place these results in context, Table~\ref{tab:ifu_compare} summarises the instrumental parameters most relevant for mapping extended Ly$\alpha$ emission with IFU spectroscopy.


\section{Future observations of Ly$\alpha$ nebulae at Cosmic Dawn: the role of SHARP's multi-IFU VESPER}
\label{sec:future}

Upcoming wide-field and deep surveys, such as those of Euclid, LSST, and JWST, are expected to reveal a growing population of quasars and faint AGN beyond $z \sim 7$, extending up to and possibly beyond $z \sim 10$ \citep[e.g.][]{Barnett2019, Trinca2023, Schneider2023}.
These sources will be prime laboratories for studying the early growth of supermassive black holes, their impact on their host galaxies, and their role in driving cosmic reionisation. However, at these redshifts, the CGM and large-scale gas flows around AGN are still almost entirely unconstrained.

To address these questions we require spatially resolved spectroscopy of the Ly$\alpha$-emitting gas in such systems, which VESPER on SHARP is ideally suited to deliver.
In particular, VESPER’s wavelength coverage ($1.2$-$2.4 \, \mu$m) and spectral resolution ($R\sim 3000$) will make it possible to observe Ly$\alpha$ at $z>9$ and to investigate in detail the kinematics of the emitting gas. Thanks to its high spatial resolution (31 mas pixel$^{-1}$) and multiplexing capability (12 IFU, each with a $\sim 1.7^{\prime\prime} \times 1.5^{\prime\prime}$ field of view, covering a total area of about $24^{\prime\prime} \times 70^{\prime\prime}$), VESPER will enable both the detection and spatially resolved characterization of extended Ly$\alpha$ emission around AGN at these redshifts, paving the way for the exploration of the circumgalactic medium in the first billion years of cosmic history (see Table~\ref{tab:ifu_compare} for a comparison with the other IFU spectrographs).

At $z>9$, deep in the epoch of reionisation, galaxies are expected to be extremely gas-rich, with star-forming regions enshrouded in dense clouds of neutral hydrogen that efficiently absorb Lyman–$\alpha$ photons \citep{Heintz2024}. At the same time, the IGM becomes increasingly neutral toward higher redshifts \citep{DeBarros2017}, leading to enhanced resonant scattering of Lyman–$\alpha$ emission. The combined effect of internal absorption and IGM scattering implies that detectable Lyman–$\alpha$ emission should be limited to the late stages of reionisation, around $z \sim 6$ \citep[e.g.,][]{Pentericci2011, Madau2024}.
Nevertheless, both simulations \citep[e.g.,][]{Witten2023, Costa2022} and recent observations \citep[e.g.,][]{Tang2023, Bunker2023, Saxena2023, Nakane2024, Chen2024, Jung2024, Napolitano2024, Witstok2025Nature} indicate that Lyman–$\alpha$ emission can still be observed at redshifts well within the reionisation era. This is likely facilitated by the presence of large ionised bubbles surrounding luminous sources, formed through local overdensities and the collective ionising output of nearby companions \citep[e.g.,][]{Tilvi2020, Leonova2022, Witstok2024, Witstok2025, Witten2023}.
A Lyman–$\alpha$ nebula has recently been detected with the NIRSpec IFU \citep{Scholtz2024} around an AGN at $z = 10.6$ (GN-z11). The SB profile of the halo closely resembles those observed around quasars at lower redshift (e.g. Farina et al. 2019), while the black hole mass and bolometric luminosity of GN-z11 ($M_{BH} \sim 10^{6} M_{\odot}$, $L_{bol} \sim 10^{45}$ erg s$^{-1}$; \cite{Maiolino2024}) are characteristic of AGN rather than quasars.
While JWST has already obtained NIRSpec/IFU observations for a small number of galaxies at $z\gtrsim 9$ \citep[e.g.][]{Messa2026,Pascalau2026}, GN-z11 currently provides the only published case of spatially extended Ly$\alpha$ emission at these redshifts in NIRSpec/IFU data \citep{Scholtz2024}. In the case of GN-z11, the deep IFU observations were primarily motivated by the study of rest-UV diagnostics (e.g. He\,\textsc{ii} and C\,\textsc{iii}]) and the nature of the ionising sources in its environment, with the Ly$\alpha$ halo reported as an additional result \citep{Maiolino2024b,Scholtz2024}. The detection of an extended Lyman–$\alpha$ emission around GN-z11 demonstrates that such nebulae can indeed be observed well into the epoch of reionisation, and confirms that VESPER will be capable of studying similar systems at $z > 9$. More importantly, it suggests that we will not need to rely on the detection of rare, extremely luminous quasars at these redshifts, but will instead be able to map Lyman–$\alpha$ nebulae around the more numerous AGN population at redshift even higher than that of GN-z11. 

More generally, spatially resolved studies of extended Ly$\alpha$ haloes at $z>9$ remain limited by the small number of suitable targets, by the strong source-to-source dependence of Ly$\alpha$ visibility during reionisation, and by the depth and PSF subtraction requirements needed to reach low surface-brightness levels. In addition, the NIRSpec/IFU field of view is $\sim3^{\prime\prime}\times3^{\prime\prime}$, which is well matched to the inner CGM but inefficient for tracing emission on tens-to-hundreds of kpc scales without mosaicking. VESPER will complement JWST by providing higher spectral resolution and multiplexed IFU mapping over a much larger area.

Understanding such systems, however, requires not only detecting the extended emission but also resolving its kinematics, as the velocity structure of the Lyman–$\alpha$ line provides essential insight into the physical processes shaping the CGM.
Studies at lower redshift have shown that Lyman–$\alpha$ kinematics can reveal signatures of feedback and large-scale gas motions. For instance, \citet{Ginolfi2018} analysed the Lyman–$\alpha$ nebula surrounding a broad absorption line (BAL) quasar at $z \sim 5$. Despite the complex line profile, their velocity-dispersion map revealed strong broadening in the inner CGM, about twice the values typically observed in radio-quiet quasars \citep[e.g.,][]{Borisova2016}. This broadening was interpreted as evidence for either Lyman–$\alpha$ emission directly tracing a large-scale outflow or reflecting the turbulence injected into the CGM by the quasar-driven wind.
Similarly, \citet{Travascio2020} studied a hyperluminous quasar at $z = 3.6$ and found that regions showing a skewed Lyman–$\alpha$ line profile (see toy model in \citealt{Verhamme2006}) coincided with the presence of blue-shifted ionised outflows traced by He\,\textsc{ii}, suggesting that the Lyman–$\alpha$ emission is linked to quasar feedback.
These are just two examples illustrating how critical it is to resolve kinematics in order to connect the energetics on large CGM scales with those on the small scales of the AGN. The higher spatial resolution provided by VESPER will enable mapping of the Lyman–$\alpha$ emission on much smaller physical scales, down to one tenth of those probed by MUSE at $z = 3.6$ \citep{Travascio2020}. 
At $z \sim 10$, VESPER’s spatial sampling of $0.031^{\prime\prime}$\,pixel$^{-1}$ corresponds to $\sim150$ pc per pixel, compared to $\sim1.5$ kpc per pixel for MUSE at $z = 3.6$. In the MUSE observations, the seeing-limited PSF implies that Lyman–$\alpha$ kinematics are effectively traced on several-kiloparsec scales. In contrast, with VESPER assisted by the MCAO we will be able to derive Lyman–$\alpha$ velocity and dispersion fields using modest $0.1^{\prime\prime}$–$0.2^{\prime\prime}$ spatial binning, corresponding to $\lesssim0.5$–$1$ kpc physical scales at $z\sim9\text{--}10$, while retaining the native sampling for detailed morphological studies of quasar-driven outflows and their interaction with the surrounding CGM.

\begin{figure*}
\includegraphics[width=1.0\textwidth]{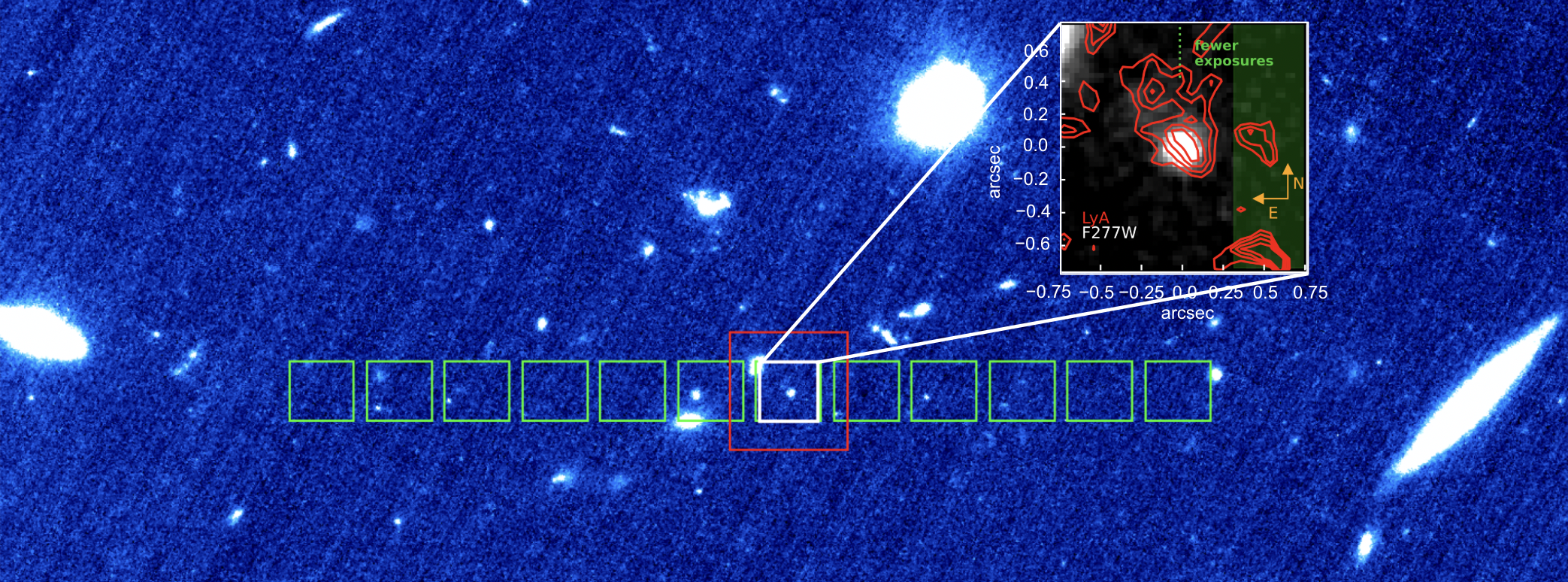}	
\caption{A comparison between the region published by \cite{Scholtz2024}, a $1.5^{\prime\prime} \times 1.5^{\prime\prime}$ JWST/NIRCam F277W continuum image overlaid with red dashed contours tracing the Ly$\alpha$ emission detected at the 2, 3, 4, and 5$\sigma$ levels, and the area covered by an array of 12 VESPER IFUs (shown in green), which spans $\sim 24^{\prime\prime}$. At $z \sim 9$, this corresponds to a physical scale of roughly $\sim 100$ kpc. For reference, the NIRSpec IFU field of view ($3^{\prime \prime} \times 3^{\prime \prime}$) used for the GN-z11 observations is shown in red.}
\label{fig:1}
\end{figure*}

\subsection{Observation strategy and expected performance of SHARP with VESPER}

The REQUIEM survey observations of quasars at $z\nobreakspace{}\sim\nobreakspace{}6.6$ with MUSE \citep{Farina2019} provide a valuable benchmark for evaluating the feasibility of similar studies at higher redshifts. 
In that work, observations were carried out down to a SB limits between $10^{-17}$ and $ 10^{-18} \mathrm{erg \, s^{-1} \,cm^{-2}\  arcsec^{-2}}$, typically reached in exposures of about one hour per source (apart from a few exceptions), yielding S/N $\sim 5$. 
We simulated an observation with VESPER’s ETC (version 0.6) aiming for S/N $\sim 5$ on a SB limit comparable to the average case in \cite{Farina2019} at $z \sim 6.6$, i.e. SB$ \sim 4.5 \times 10^{-18}$ erg s$^{-1}$ cm$^{-2}$ arcsec$^{-2}$. After accounting for the cosmological dimming between $z = 6.6$ and $z = 9$, we scaled this limit by a factor of $0.33$, obtaining a target SB limit of $1.5\times10^{-18}$ erg s$^{-1}$ cm$^{-2}$ arcsec$^{-2}$. For the line profile, we adopted a FWHM of $4$ \AA\ ($\sim 100$ km s$^{-1}$), corresponding to 3-4 spectral channels for VESPER. This width represents a typical narrow Ly$\alpha$ component arising from individual CGM gas clumps within the nebula. The central wavelength was set to $12160$ \AA\, corresponding to Ly$\alpha$ at $z \sim 9$. Integrating the flux over $1$ arcsec$^{2}$ and over the 3-4 spectral channels covering the line, the ETC predicts a S/N $\sim 5$ for a total exposure of $4$ hours, consistent with simple scaling arguments based on Farina’s measurements.

At these depths, Ly$\alpha$ emission in the central $\lesssim10$–$20$ kpc of quasar halos is expected to reach SB levels of a few $\times10^{-18}$ to $10^{-17}$ erg s$^{-1}$ cm$^{-2}$ arcsec$^{-2}$, i.e. well above the detection threshold. Under these conditions, it will be possible not only to detect the halos, but also to extract reliable Ly$\alpha$ line profiles in sub-kiloparsec apertures across much of the nebula. This corresponds to a moderate spatial binning of $0.1^{\prime\prime}$-$0.2^{\prime\prime}$ and will enable measurements of velocity gradients and line-width variations within the CGM. At the same time, the native $0.031^{\prime\prime}$ sampling ($\sim150$ pc at $z\sim9$=$10$) will preserve the fine morphological substructure of the emission, allowing us to resolve individual clumps and filaments in the Ly$\alpha$ nebula and to directly assess the clumpiness of the CGM at these redshifts. Compared to the few-kiloparsec scales accessible with current MUSE observations, this will allow a much more detailed view of the high-redshift CGM.

At the redshifts accessible to Lyman–$\alpha$ with SHARP, a single VESPER IFU will cover a region of approximately $8 \times 7$ kpc$^{2}$. Given the expected sizes of Lyman–$\alpha$ nebulae at these redshifts \citep{Scholtz2024}, one VESPER IFU would already be sufficient to encompass the entire region around a quasar at $z\sim 9$. However, mosaicking multiple IFUs would enable the exploration of more extended, low SB structures, as might be expected in dense environments. By exploiting VESPER’s multiplexing capability, several IFUs could be deployed simultaneously to map areas up to $\sim 100$ kpc, comparable to the typical extent of the CGM  (see a comparison of the field of GN-z11 covered by JWST and by an array of 12 VESPER's IFU in Fig. \ref{fig:1}). 
Such an approach would allow the simultaneous investigation of large-scale gas distributions and small-scale outflow structures, tracing complementary aspects of baryon and metal cycling around early quasars.

Finally, the optimal sample size will be driven by the number of confirmed $z>9$ AGN/quasars delivered by forthcoming surveys and by the observing time required to reach the low surface-brightness regime. Even a single well-characterised detection at $z>9$ would already provide a decisive proof of feasibility and a benchmark for models of CGM structure during reionisation. However, addressing the broader questions outlined in the first sections (e.g. diversity of CGM morphology/kinematics, connection to AGN properties and environment, and the incidence of extended emission) will require a small statistical sample. A practical goal is a two-tier programme: (i) deep observations of a few benchmark systems ($\sim$2-3) to obtain high-fidelity maps and kinematics, followed by (ii) observations of a sample of order $\sim$10, as suitable targets become available, to quantify detection fractions, characteristic sizes, and the range of kinematic behaviours. Such a programme can naturally scale with the availability of targets.

Besides Ly$\alpha$, VESPER’s wavelength coverage will also enable IFU studies of extended nebulae in other key diagnostic lines, such as C\,\textsc{iv} and He\,\textsc{ii} at $z\sim6.8$ and $z\sim6.3$, respectively, for REQUIEM-like quasars, tracing metal-enriched, highly ionised gas in the CGM and potentially revealing signatures of quasar-driven outflows. Spatially resolved maps of metal-line ratios will further help disentangle ionization conditions from metal enrichment and constrain the physical state of the CGM. 
In addition, H$\alpha$ will be observable with VESPER over $0.8 \lesssim z \lesssim 2.6$, providing a non-resonant tracer of the ionised gas distribution and kinematics that is less affected by resonant scattering than Ly$\alpha$. Combining H$\alpha$ observations with VESPER and Ly$\alpha$ observations with KWCI would therefore offer powerful, complementary constraints on the physical conditions and radiative-transfer effects in the CGM.

\section{Summary and conclusions}

Extended Ly$\alpha$ nebulae are a powerful probe of the CGM and of the coupling between AGN, their host galaxies, and the surrounding environment. Over the past decade, integral field spectroscopy with instruments such as MUSE and KCWI has revealed that extended Ly$\alpha$ emission is nearly ubiquitous around quasars up to $z\sim6$, highlighting the key role of the CGM in galaxy and black hole co-evolution.
Recent JWST observations have pushed these studies to higher redshifts, into the epoch of reionisation, although their limited spectral resolution still hampers detailed investigations of the gas kinematics. SHARP at ELT, with its multi-IFU module VESPER, will build upon these advances, enabling spatially resolved studies of the CGM around AGN at z>9, within the first few hundred million years of cosmic history, and providing unprecedented insights into the processes governing gas accretion, feedback, and the early growth of supermassive black holes and their host galaxies.
A tiered programme, combining a few deep benchmark targets with a modest sample as new $z>9$ AGN are discovered, would provide both proof of feasibility and the first statistical view of CGM structure and kinematics at these epochs.

\printcredits

\vspace{0.5cm}


\bibliographystyle{cas-model2-names}

\bibliography{bib}


\end{document}